# Mechanical deformation of monocytic THP-1 cells : occurrence of two seqential phases with differential sensitivity to metabolic inhibitors

**Fabienne Richelme, Anne Marie Benoliel and Pierre Bongrand**

Laboratoire d'Immunologie, INSERM U 387, Hôpital de Sainte-Marguerite, BP 29, 13274 Marseille Cedex 09 FRANCE[1]

**Blood leukocytes can exhibit extensive morphological changes during their passage through small capillary vessels. The human monocytic THP-1 cell line was used to explore the metabolic dependence of these shape changes. Cells were aspirated into micropipettes for determination of the rate of protrusion formation. They were then released and the kinetics of morphological recovery was studied. Results were consistent with Evans' model (Blood, 64 : 1028, 1984) of a viscous liquid droplet surrounded by a tensile membrane. The estimated values of cytoplasmic viscosity and membrane tension were 162 Pa.s and 0.0142 millinewton/m respectively. The influence of metabolic inhibitors on cell mechanical behaviour was then studied : results strongly suggested that deformation involved two sequential phases. The cell elongation rate measured during the first 30 seconds following the onset of aspiration was unaffected by azide, an inhibitor of energy production, and it was about doubled by cytochalasin D, a microfilament inhibitor, and colchicine, a microtubule inhibitor. However, during the following two minutes, deformation was almost abolished in cells treated with azide and cytochalasin D, whereas the protrusion of control cells exhibited about threefold length increase.**
**It is concluded that, although cells seemed to deform as passive objects, active metabolic processes were required to allow extensive morphological changes triggered by external forces.**

Leukocyte functions are heavily dependent on the cells' capacity to undergo extensive deformation following mechanical or biochemical stimulation. Thus, passage through narrow capillary vessels requires that blood leukocytes transiently acquire an elongated sausage shape (Bagge & Branemark, 1977 ; Worthen et al., 1989). Passage between endothelial cells during diapedesis necessitates extensive squeezing through a narrow aperture (Marchesi, 1961). Migration on a surface involves acquisition of a typical hand mirror shape, with continuous emission of forward lamellipodia and rearward contraction (Zigmond, 1978). Adhesion may also involve extensive deformation with concomitant spreading at the micrometer level and smoothing of microvilli on the submicometer scale (Mège et al., 1987).

It is therefore of high interest to understand the leukocyte response to mechanical stimulation. This may yield direct information on passive cell behaviour in flowing blood. Also, this knowledge is a prerequisite to a thorough understanding of the mechanisms of shape control during active deformation, e.g. during migration through the surrounding environment. Several procedures were devised to study cell mechanical properties, including analysis of deformation induced by centrifugal forces (Landau et al., 1954), measurement of the indentation produced by a solid rod applied on the cell surface with known force (Petersen et al., 1982), study of the displacement of intracellular magnetic particles subjected to exogenous forces (Valberg and Albertini, 1985), Twisting particles attached to surface receptors by means of a magnetic field (Wang et al., 1993) or aspirating cells into glass micropipettes with controlled pressure and subsequent monitoring of induced protrusion (Mitchison and Swann, 1954). The latter procedure was extensively used to study the mechanical properties of blood neutrophils. Small deformations were described with a standard viscoelastic model (Schmid-Schönbein et al., 1981). Larger shape changes, involving the formation of a protrusion substantially longer than the pipette radius, were accounted for by modelling neutrophils as viscous liquid droplets surrounded by a membrane under tension (Evans & Kukan, 1984 ; Evans & Yeung, 1989 ; Yeung & Evans, 1989). The cytoplasmic viscosity was on the order of 200 Pa.second at room temperature, and the membrane tension was about 0.035 millinewton/m.

An important question is to know whether cells subjected to micropipette aspiration may be considered as passive objects with constant properties and to what extent material parameters are dependent on time and metabolic properties. First, micropipette aspiration may indeed trigger active cell responses. Thus, Evans and Kukan (1984) observed that neutrophils might display erratic movements a few minutes after the onset of aspiration. Further, micropipette-aspirated RBL cells (Horoyan et al., 1990) or neutrophils (Zaffran et al., 1993) often exhibited transient increases of cytosolic calcium concentration and reorganisation of actin microfilaments (Horoyan et al., 1990). Second, metabolic inhibitors were reported to alter cell mechanical behavior : thus, glycolytic inhibitors such as fluoride, iodoacetate or deoxyglucose reduced the deformability of blood polymorphonuclear leukocytes (Miller and Myers, 1975). A temperature decrease also reduced deformation (Sung et al., 1982). Finally, treating leukocytes with cytochalasin B, a microfilament inhibitor, increased deformability (Mège et al., 1985 ; Tsai et al., 1994). Similar results were obtained on fibroblasts (Erickson, 1980). However, cytochalasin inhibited the formation of protrusion by micropipette-aspirated *Fundulus heroclitus* egg cells (Tickle and Trinkaus, 1977). Finally, activating neutrophils with chemotactic factors increased deformability at low concentration, and increased rigidity at higher concentration (Kawaoka et al., 1981).

Thus, it seemed useful to obtain more detailed information on the influence of metabolic inhibitors on leukocyte response to mechanical forces. In the present work, human monocytic THP-1 cells were subjected to micropipette aspiration under control conditions or after being treated with colchicine (a microtubule inhibitor), cytochalasin D (a microfilament inhibitor) or azide (an inhibitor of electron transport). The kinetics of protrusion formation was monitored : during the first 30 seconds following the onset of aspiration, protrusion length linearly increased with respect to time, and this increase was enhanced by colchicine and cytochalasin D, whereas it was not significantly affected by azide. However, during the following two minutes, both azide and cytochalasin D reduced deformation rate. It is concluded that mechanically induced cell deformation involved two sequential phases with differential dependence on the cell metabolic status.

**MATERIALS AND METHODS**

**Cells and media**. The human monocytic THP-1 cell line (Tushiya et al., 1980) was kindly donated by Dr. F. Birg (INSERM U 119, Marseille). Cells actively ingested opsonized erythrocytes and displayed typical antigens of the mononuclear phagocyte lineage, including CD11b, CD18, CD35, and CD64, and non polymorphic epitopes of class I histocompatibility molecules, as checked with flow cytometry (not shown). However, they did not adhere strongly to glass surfaces. They were maintained in RPMI 1640 medium (Flow laboratories, McLean, VA) supplemented with 20 mM HEPES, 10 % fetal calf serum (Flow), 2 mM L-glutamine, 50 U/ml penicillin and 50 µg/ml streptomycin. Aspiration was performed in saline supplemented with 20 mM HEPES, 5 mM KCl, 1 mM $MgCl_2$, 1 mM $CaCl_2$, 10 mM glucose and 1 mg/ml bovine albumin.

**Cell labeling**. Cytoplasmic labeling was performed as previously described (André et al., 1990) by incubating cells for 15 minutes at 37°C in culture medium supplemented with 40 µg/ml fluorescein diacetate (Sigma). Polymerized actin was revealed (Horoyan et al., 1990) by fixing cells on the tip of the micropipette for 20 minutes with 3.7 % paraformaldehyde, and then labeling samples by another 20 minute incubation in phosphate buffer containing 10 U/ml bodipy phallacidin (Molecular Probes, Eugen, OR) and 0.1 mg/ml lysophosphatidylcholine (Sigma).

**Metabolic inhibitors**. In some experiments, cells were preincubated for 30 minutes at 37°C in 10 µg/ml cytochalasin D (Sigma, St Louis, Mo) to inhibit microfilament assembly. Alternatively, microtubules were inhibited by 30 minute incubation with 0.1 mM colchicine (Sigma). Cytochrome function was inhibited by 30 minute incubation with 10 mM sodium azide.

**Micropipettes**. Micropipettes were obtained from borosilicate capillary tubes of 0.8 mm internal diameter (Clark electromedical instruments, provided by Phymep, Paris), using a programmable horizontal mipropipette puller (Campden instruments, model 773). In some cases, the aperture was enlarged by breaking the pipette tip with a deFonbrune microforge (Alcatel, Paris). The inner diameter of the pipette tip was set at 5.5 µm, i.e. about one third of THP-1 cell average diameter.

**Microscope and aspiration stage**. The micropipette was connected to a pressure generator composed of a U-tube connected to a syringe mounted on a syringe holder. The pressure was measured with a sensor (ref. 78300-04, Cole Parmer instruments) connected to an analog input card (model PC-ADC 12B8V/D, Digimetrix, Perpignan, France) mounted on an IBM compatible desk computer. This allowed real time pressure determination with about 1 Pa resolution.

Aspiration was performed on the stage of an Olympus IMT2 inverted microscope bearing a high sensitivity SIT videocamera (Lhesa model 4036, Cergy Pontoise, France). The output was connected to a PCVision+ digitizer (Imaging Technology, Bedford, MA) mounted on the computer. The digitized image was subjected to reverse digital-to-analog conversion and displayed on a monitor with continous video recording. This procedure allowed continous superimposition of time and pressure on video images. A representative image is shown on Figure 1. Also, when fluorescence labeling was performed, the amount of fluorescence in the cell body and protrusion was determined as previously described (Horoyan et al., 1990).

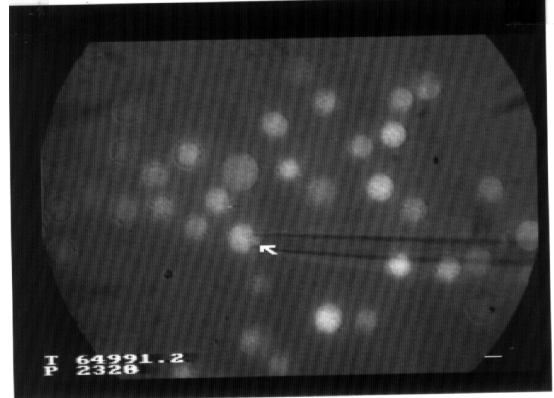

**Figure 1. Cell aspiration into a micropipette**. Human monocytic THP-1 cells were labeled with Fluo-3, a fluorescent calcium probe. A cell was then was aspirated into a micropipette with a pressure of about 98 Pa under continuous video recording. The microscope field was illuminated with a combination of visible and UV light. The protrusion is clearly visible (arrow). Time T is indicated (with 0.1 second accuracy) together with pressure P (with units of about 1.1 Pa). Bar length is 10 micrometers.

**Experimental procedure**. In a typical experiment, a 300 µl aliquot of cell suspension ($4 \times 10^5$/ml) was deposited on a glass coverslip (20x60 mm$^2$) set on the microscope stage. After zero pressure determination, a series of typically 10-20 cells were aspirated with a negative pressure of about 100 Pa (~1 cm $H_2O$) for 150 seconds each. The whole experiment was recorded for later analysis. The kinetics of protrusion formation and elongation was then studied by following the motion of the protrusion with a dedicated software evolved in the laboratory (Kaplanski et al., 1993). Pixel size was about 0.15 µm.

**Data analysis**.
*Model for extensive deformation*. Following Evans and Kukan (1984), cells were considered as viscous liquid droplets (of viscosity µ) surrounded by an elastic membrane with constant tension T. Since some adhesion persisted with all tested protein concentrations (not shown), a fairly low protein concentration was used and a model

accounting for adhesion was elaborated as follows (Figure 2)

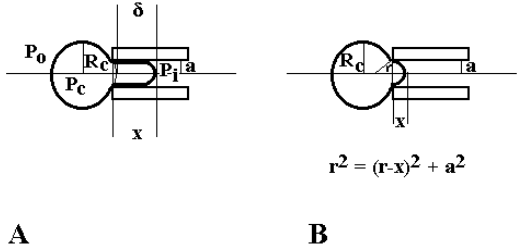

**Figure 2. Geometrical model for protrusion formation.** Cells are modelled as spheres of constant radius $R_c$ with a cylindrical protrusion terminated with an hemispherical tip of radius a equal to the inner pipette radius (Fig. 2 A). Parameter x is the distance between the pipette entry and the protrusion tip. Parameter $\delta$ is the distance between the protrusion tip and the cell boundary before deformation. The pressure is $P_i$ within the pipette, $P_c$ within the cell and $P_0$ in the extracellular medium. Fig. 2 B shows the cell response to subthreshold pressure : the protrusion is a spherical cap with radius r comprised between a and $R_c$.

Since fairly large pipettes were used, the pressure dissipation within the cell cortex was deemed negligible, in accordance with Evans and Yeung's findings (1989). The pressure dissipation during cell deformation was thus written as the sum of two terms :
- The pressure drop upstream from the pipette mouth was approximated as the exact value calculated for a fluid flowing through a circular orifice bounded by a no-slip plane (Happel & Brenner, 1991 - See Fig. 6 of Yeung & Evans, 1989), yielding :

$$P_1 = 3Q\mu/a^3 \qquad (1)$$

where Q is the flow rate and a is the pipette radius.
- Now, the pressure decrease within the protrusion was calculated assuming zero velocity near the pipette wall. Thus, following Poiseuille's law, we used the following formula:

$$P_2 = 8\mu LQ/\pi a^4 \qquad (2)$$

where L is the protrusion length (Figure 2A). By combining Eqs (1) and (2), and replacing Q with $(dL/dt)/\pi a^2$, The following analytical formula was obtained :

$$L(t) = (-3\pi a/8 + (9\pi^2 a^2/64 + a^2 Pt/\mu)^{(1/2)})/2 \qquad (3)$$

Laplace law was used to relate the driving pressure P to the pressure difference $P_0-P_i$ between the external medium and the pipette interior (Figure 2A) :

$$P = (P_0-P_i) - 2T(1/a - 1/R_c) \qquad (4)$$

where $R_c$ is the cell radius (this is fairly constant during aspiration under our experimental conditions). The assumption of constant membrane tension T may be reconciled with the assumption of no relative displacement between the cell surface and the pipette wall if anchoring is mediated by tranmembrane proteins directely connected to the protrusion core.

*Direct determination of membrane tension.* We attempted to achieve direct determination of the membrane tension by subjecting aspirated cells to small sequential pressure increases (of about 20 Pa each) in order to determine the minimal pressure required to generate a fairly spherical protrusion. As shown on Figure 2B, the displacement $\delta$ of the protrusion tip generated by indefinitely slow aspiration with a low pressure P0-Pi is given by the following equations :

$$r = 1/(1/R_c + (P_0-P_i)/2T) \qquad (5)$$

(this is Laplace law relating the radius of curvature of the protrusion tip and sucking pressure)

$$\delta = r - (r^2 - a^2) - R_c + x - (R_c^2 - a^2)^{1/2} \qquad (6)$$

(as obtained with elementary geometrical calculations)
Thus, the membrane tension T could be obtained by numerical fit of the variations of $\delta$ versus $P_i$ with a theoretical curve derived from equations (5) and (6).

*Use of relaxation experiments to derive the ratio between surface tension and core viscosity.* A general argument supporting Evans' model is the general finding that micropipette-aspirated cells recover a spherical shape when they are released from the pipette. Further, Tran-Son-Tay et al. (1991) provided a simple numerical equation allowing easy determination of the ratio $T/\mu$ between cell tension and cytoplasmic viscosity by plotting the variations of the highest (L) and lowest (W) cell diameter versus nondimensional time $t_{nd} = tT/\mu R_c$, where t is time. Thus, some cells were deformed into elongated "sausage" shape for about ten seconds, then released under video recording to determine the kinetics of shape recovery.

## RESULTS

**Relaxation studies**.
In a first series of experiments, cells were aspirated into large micropipettes in order to induce a "sausage-like" shape. They were released after a 10-20 second delay, under continuous video recording. First, it was checked that cells recovered a spherical shape, in accordance with the viscous droplet model. Second, as shown on Figure 3, the experimental plot of variation of the cell highest and lowest diameter versus time was used to derive the ratio between membrane tension (T) and interior viscosity (μ), as described by Tran-Son-Tay et al. (1991). This yielded a best fit value of 0.088 μm/second

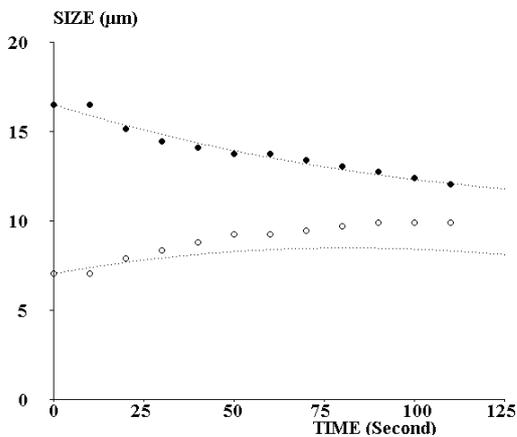

**FIGURE 3 : MORPHOLOGICAL RECOVERY AFTER DEFORMATION**. A THP-1 cell was aspirated into a pipette until it acquired a "sausage" shape with length L and width W. After a 10 second wait, the cell was expelled and parameters L (closed circles) and W (open circles) were measured at regular intervals. The dotted line represents a theoretical curve calculated with a numerical formula provided by Tran-Son-Tay et al. (1991). The ratio between tension and viscosity is $8.8 \times 10^{-8}$ m/s.

**Cell deformation is consistent with a model assuming constant membrane tension and cytoplasmic viscosity together with cell-to-glass adhesion.**

Twenty two control cells (mean radius : 8.6 μm) were aspirated under constant pressure. The time dependence of protrusion length is shown on Figure 4. Equations (3) and (4) were then used together with the experimental value of T/μ : the viscosity was derived by least square error minimization. Clearly, the fitted theoretical curve displayed satisfactory agreement with experimental data. Note also that deformation remained moderate, since the protrusion did not include more than about 10 % of the total cell volume, thus allowing to neglect the variations of the radius of the spherical part of these cells.

The above procedure yielded numerical estimates of 162 Pa.second and 0.0142 mN/m for the cytoplasmic viscosity and membrane tension respectively.

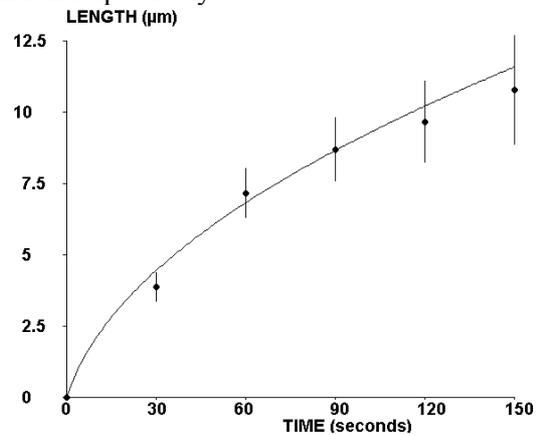

**Figure 4 : Kinetics of protrusion formation in a micropipette**. Twenty-two THP-1 cells were aspirated into a micropipette with a pressur of 98 Pa. Average values of the protrusion length at time 30, 60, 90, 120 and 150 s are shown (vertical bar length is twice the standard error) together with fitted theoretical curve obtained by fitting the viscosity and using the tension/viscosity radio derived from relaxation studies.

**Direct determination of membrane tension**.
Seven cells were aspirated with sequential steps of small enough pressure increases (about 30 Pa) that no extensive protrusion could appear. Also, pressure increases were separated by time intervals of several tens of seconds in order that equilibrium deformation might be reached. Experimental data are shown on Figure 5 together with fitted theoretical curves, yielding a membrane tension of 0.0914 mN/m.

**Effect of metabolic inhibitors.**
Although a two-parameter model did not quantitatively account for all aspects of THP-1

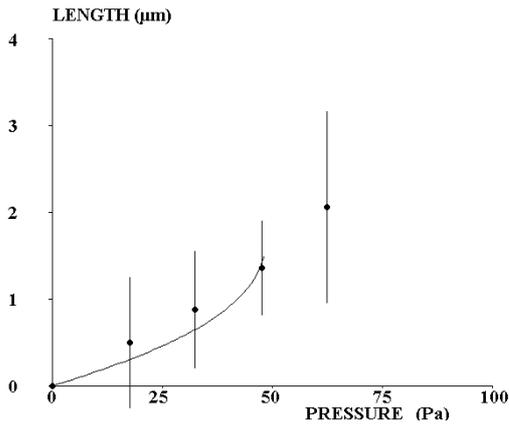

**Figure 5 : Initial cell deformation induced by subthreshold pressure.** Seven THP-1 cells were aspirated with increasing pressure remaining too low to induce the formation of a long protrusion. The displacement of the cell boundary was determined 20-30 secondes after a pressure change. Mean displacements are shown (vertical bar length is twice the standard error) together with a theoretical curve calculated by modeling the cell surface as an elastic membrane with constant tension.

cell behaviour during aspiration, we obtained a qualitative fit between experimental sets of data and theoretical curves (Figures 3-5), making cells appear as passive objects. It was of interest to know how metabolic inhibitors might alter experimental parameters. Therefore, aspiration experiments were repeated on cells treated with inhibitors of cytoskeletal assembly or energy production. Results are summarized on Table 1. A striking conclusion was that these inhibitors differentially affected cell behaviour at different periods of time following the onset of aspiration. Indeed, during the first 30 seconds, protrusion elongation rate was not affected by azide but it was significantly ($P<0.001$) and substantially (about twofold) increased by colchicine or cytochalasin D. However, during the following 120 seconds, the deformation of azide- and cytochalasin-treated cells was nearly blocked, whether the length of the protrusion yielded by control or colchicine-treated cells displayed about threefold increase, strongly suggesting that different mechanisms were involved in the early and late phase of cell deformation.

**Cytoskeletal behaviour during cell deformation**. It was important to know whether cytoskeletal components remained bound to the cell membrane during deformation. This question was addressed quantitatively by comparing the distribution of cytoplasmic and cytoskeletal markers in micropipette-aspirated cells. First, cells were labeled with fluorescein diacetate and aspirated for 3 minutes with a pressure of 100 Pa. They were then studied with fluorescence microscopy and image analysis. The mean ratio between the total fluorescence of the cell body and the protrusion was $2.41 \pm 0.10$ (standard error of the mean, n = 27 cells). Polymerized actin was then labeled on individual cells. The mean ratio between the total fluorescence of the cell body and protrusion was $2.52 \pm 0.22$ (standard error of the mean, n = 14). Therefore, the density of polymerized actin was comparable in the cell body and the protrusion. However, when individual micrographs were observed, very limited fluorescence accumulation was sometimes apparent in the protrusion. A typical image is displayed on Figure 6.

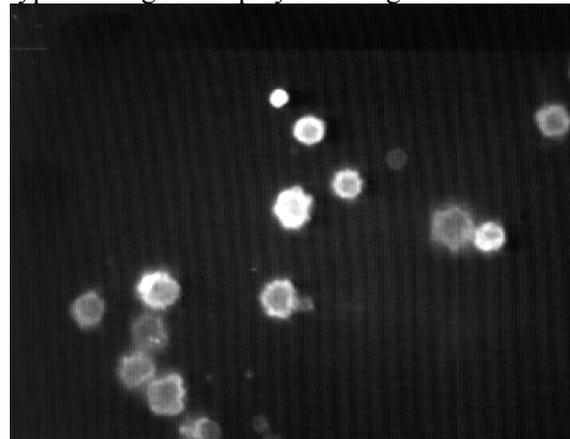

**Figure 6 : Microfilament behaviour during cell deformation**. Fourteen THP-1 cells were aspirated for 3 minutes in a micropipette with a pressure of 100 Pa. They were fixed and stained with bodipy phallacidin, and then examined with fluorescence microscopy. A typical image is shown : The fluorescence is fairly homogeneous in the protrusion, but a bright fluorescent point is visible on the tip.

**DISCUSSION**
The aim of this work was twofold. First, it was necessary to characterize the behaviour of THP-1 cells subjected to micropipette aspiration. Second, we wished to analyze the effect of metabolic inhibitors on these properties.
Three main procedures were used to study cell mechanical properties : first, we studied the cell relaxation properties after at least ten

### TABLE 1

### INFLUENCE OF METABOLIC INHIBITORS ON CELL DEFORMATION

| | Protrusion length (micrometer) | | | | |
|---|---|---|---|---|---|
| Time (second) | 30 | 60 | 90 | 120 | 150 |
| Control (n=22) | 3.87 (±0.51) | 7.17 (±0.85) | 8.70(±1.12) | 9.67(±1.41) | 10.78(±1.90) |
| Cytochalasin D (n=28) | 7.26(±0.47) P<0.001 | 6.96(±0.33) | 7.70(±0.41) | 7.73(±0.58) | 7.69(±0.38) |
| Colchicine (n=24) | 6.30(±0.23) P<0.001 | 8.01(±0.34) | 9.49(±0.51) | 10.08(±0.88) | 10.21(±0.52) |
| Azide (n=20) | 3.28(±0.31) | 3.82(±0.29) P<0.001 | 3.88(±0.33) P<0.001) | ND | ND |

Monocytic THP-1 cells were aspirated into micropipettes with or without being treated with metabolic inhibitors and the length of induced protrusions was measured at regular intervals. Average values obtained on a number (n) of cells ranging between 20 and 28 are shown ± standard error of the mean. The significance of effect of a given treatment was calculated with Student's t test.

second deformation in a micropipette. Second, we studied the kinetics of extensive deformation. Third, we studied the small deformation induced by **prolonged** exposure to weak forces, These conditions were devised to avoid the influence of short-term elastic behaviour (Schmid-Schönbein et al., 1981 ; Tran-Son-Tay et al., 1991), thus allowing a safer use of Evans' model of a viscous liquid droplet surrounded by a tensile membrane. Qualitative agreement was found between experimental data and theoretical predictions (Figures 3, 4 & 5) and the experimental estimates of 0.014 millinewton/m and 162 Pascal-second for tension and viscosity compare well to the neutrophil values (i.e. 0.035 mN/m and ~ 200 Pa.s, Evans & Kukan, 1984, Evans & Yeung, 1989 or 0.024 mN/m and 150 Pa.s, Needham & Hochmuth, 1992 ; Tran-Son-Tay et al., 1991). Comparable values were also reported for the cytoplasmic viscosity of HL-60 promyelocytic cells (Tsai et al., 1996). The quantitative discrepancy between the tension values determined with large and small deformations might also be ascribed to either non-newtonian behaviour of the cell interior (Tsai et al., 1993) or substantial cortex elasticity (Schmid-Schönbein et al., 1981). However, the main problem with our results might be the evidence of significant cell-to-glass adhesion. The qualitative agreement between theoretical and experimental data (Figure 4) is consistent with the possibility that adhesion be mediated by a population of cell membrane proteins likely to anchor the submembranar matrix to the glass with sufficient bilayer freedom to ensure that tension be similar around the cell body and near the protrusion tip. This would warrant our modelling including constant surface tension combined with Poiseuille flow within the protrusion, however, more experiments are required to clear this point. In any case, it must be emphasized that only the simultaneous use of several complementary methods of quantifying cell deformability might help clarify aforementioned problems.

Experiments done with metabolic inhibitors suggest somewhat unexpected conclusions : comparison between experimental and theoretical data shown on Figure 4 were consistent with the view that cells behave as passive bodies with constant material parameters during aspiration. This interpretation was consistent with the finding that azide, an inhibitor of energy production, did not affect protrusion elongation rate during the first 30 secondes following the onset of aspiration. The higher deformability of cytochalasin-treated cells might be ascribed to a weakening of cytoskeletal elements, in accordance with other experimental approaches (Petersen et al., 1982 ; Wang et al., 1993). It is not surprising that azide decreased cell deformability during the second phase of aspiration. Indeed, azide was previously reported to decrease macrophage spreading in vitro (Rabinovitch and Destefano, 1974), and this inhibition was antagonized by glucose, suggesting that it might be related to a decrease of energy production. Further, the deformability of blood polymorphonuclear leucocytes was decreased by glycolytic inhibitors (Miller and Myers, 1975).

However, the clearcut finding cytochalasin also decreased cell deformability during the second phase of aspiration was somewhat unexpected. The simplest interpretation of these findings would be that some active cell participation might be required to allow cells to display a substantial level of deformation. Thus, THP-1 cells might be designed to yield to exogenous mechanical forces, which might obviously facilitate their passage through small capillary vessels. An attractive possibility would be that this yield might be triggered by a rise in intracellular calcium, since this rise was described in different populations of myeloid cells (Hoyoyan et al., 1990 ; Zaffran et al., 1993), including THP-1 cells (not shown) subjected to micropipette aspiration. However, more experiments are needed to clarify this point, since no direct link could be demonstrated between intracellular calcium changes and mechanical responses. It is planned to address this problem by studying the effect of metabolic inhibitors and intracellular calcium manipulation on phenomenological membrane tension and cytoplasmic viscosity, as determined by aspiration and relaxation studies.